\documentclass[12pt]{article}

\usepackage{amsmath}

\newcommand{\add}{\addtocounter{eqncnt}{1}}
\newcounter{eqncnt}[section]

\newcommand{\be}{\begin{equation}}
\newcommand{\ee}{\end{equation}\add}
\newcommand{\bea}{\begin{eqnarray}}
\newcommand{\eea}{\end{eqnarray}}

\newcommand{\bra}{\left\langle}
\newcommand{\ket}{\right\rangle}

\newcommand{\frw}{FLRW\ }

\begin{document}
\begin{center}
{\bf Unimodular Gravity and Averaging } \\[2mm]

\vskip .5in

{\sc A. Coley, J. Brannlund [*] and J. Latta}\\
{\it Department of Mathematics and Statistics}\\
{\it Dalhousie University, Halifax, NS B3H 3J5, Canada}

{\it aac,johanb,lattaj@mathstat.dal.ca }

{\it [*]Current address: }\\
{\it Department of Mathematics, Physics and Geology}\\
{\it Cape Breton University, 1250 Grand Lake Road}\\
{\it Sydney, NS B1P 6L2}\\
\end{center}

\begin{abstract}

The question of the averaging of inhomogeneous spacetimes in cosmology
is important for the correct interpretation of cosmological data.
In this paper we suggest a conceptually simpler approach to averaging
in cosmology based on the averaging of scalars within unimodular gravity. 
As an illustration, we consider the example of
an exact spherically
symmetric dust model, and show that within this approach averaging 
introduces correlations (corrections) to the effective dynamical evolution
equation in the form of a spatial curvature term.

\end{abstract}

\noindent [PACS: 98.80.Jk,04.50.+h] \vskip1pc

\section{Introduction}

The Universe is not isotropic or spatially homogeneous on local
scales. The correct governing equations on cosmological scales are
obtained by averaging the gravitational field equations (FE).
An averaging of inhomogeneous spacetimes in Einstein's general relativity
(GR) can lead to
dynamical behavior different from the spatially homogeneous and
isotropic Friedmann-Lema\^{i}tre-Robertson-Walker (FLRW) model; in particular, 
the expansion rate may be
significantly affected \cite{Ellis:1984}. 
Consequently, a
solution of the averaging problem is of considerable importance
for the correct interpretation of cosmological data.

The solution to this problem  necessitates a method for covariantly
(and gauge invariantly) averaging tensors on a background differential
manifold.
Unfortunately, this is a very difficult problem.
In the Isaacson  spacetime  averaging scheme in  macroscopic gravity (MG) bilocal
averaging  operators are utilized \cite{Zala}.
Choosing  a  compact  region  $\Sigma  \subset  {\cal  M}$ in an
($n$-dimensional  differentiable)  manifold  $({\cal  M}$,
$g_{\alpha  \beta })$ with a volume  $n$-form  and a  supporting
point  $x\in  \Sigma  $ to  which  the  average  value  will  be
prescribed,  the average value of a geometric  object,  $p_\beta
^\alpha  (x)$, over a region $\Sigma $ (with volume $V_\Sigma $) at 
$x\in  \Sigma $, is  defined  in  terms of the
bilocal   extension  of  the  object   $p_\beta   ^\alpha  (x)$,
$\label{bilocext}{\bf  p}_\beta ^\alpha (x,x^{\prime })=$ ${\cal
W}_{\mu  ^{\prime  }}^\alpha   (x,x^{\prime   })p_{\nu  ^{\prime
}}^{\mu  ^{\prime  }}(x^{\prime }) {\cal W}_\beta ^{\nu ^{\prime
}}(x^{\prime  },x)$, by means of the bilocal  averaging  operator
${\cal  W}_{\beta  ^{\prime  }}^\alpha  (x,x^{\prime })$.  The
averaging  scheme is covariant and linear by  construction,  and
the averaged object has the same
tensorial character as $ p_\beta ^\alpha $. In any 
manifold with a volume $n$-form
there always exist locally volume-preserving divergence-free
operators \cite{Zala},
in which the
bilocal operator ${\cal W}_\beta ^{\alpha ^{\prime }}(x^{\prime
},x)$ takes the simplest possible form: 
${\cal W}_\beta ^{\alpha ^{\prime }}(x^{\prime },x)=$  $\delta
_i^{\alpha ^{\prime }}\delta _\beta ^i$ \cite {MZ}.

The definition of an average  consequently takes on a
particularly simple form when written in 
a volume-preserving  (system of) coordinates (VPC).
Indeed, if the manifold is a pseudo-Riemannian spacetime, 
the spacetime
average of a tensor field $p_\beta ^\alpha (x),\,x\in {\cal E}$, 
at a supporting point $(t,x^a)\in {\cal E}$ in VPC is thus
\begin{equation}
\label{defaver:ED} \bra p_\beta ^\alpha (t,x^a)\ket _{{\cal
E}}=\frac 1{V_\Sigma
}\int_\Sigma 
p_\beta ^\alpha
(t+t^{\prime },x^a+x^{a\prime })dt^{\prime }d^3x^{\prime }\ .
\end{equation}

In  the MG covariant approach to the averaging problem 
the Einstein FE (EFE) on cosmological scales with a continuous distribution of cosmological matter
are modified by appropriate gravitational 
correlation (correction) terms \cite{Zala}. 
The averaged FE can always be written in the form of the FE for the
macroscopic metric tensor when the correlation terms are moved to the right-hand side of the averaged
field
equations, and consequently can be regarded as a geometric modification to the averaged (macroscopic) matter
energy-momentum tensor \cite{Zala}.
In \cite{CPZ} it was found that  by solving the MG equations
the averaged EFE for a spatially homogeneous, isotropic macroscopic spacetime
geometry has the form of the EFE of GR for an FLRW geometry with an additional spatial
curvature term (i.e., the correlation tensor is of the form of a
spatial curvature term) (see also \cite{CP}).  
Unfortunately, the  spacetime  averaging scheme in MG is
very difficult to apply and is fraught with complications \cite{Brann}.
In this paper we want to suggest an alternative approach to averaging,
exploiting the preferred nature of VPC and
based on the averaging of scalars \cite{Scalar,buch}.

\section{Unimodular gravity}

The fundamental variables  
in the action for unimodular gravity and the Einstein-Hilbert action for
GR are different  \cite{Wei89,NG99,unruh_uni}. In unimodular gravity,
there is an additional restriction on the metric, not present in GR: 
the determinant of $g_{\mu \nu}$ equals one. 
As a consequence of  $\det g_{\mu \nu}=1$, unimodular gravity is only invariant
under {\it volume-preserving} diffeomorphisms.~{\footnote{Coordinate invariance can always be
reinstated into the theory.}} Thus, unimodular gravity
presents a natural theory in which to do averaging.

Varying the action in unimodular gravity leads to the FE relating the traceless Ricci tensor,
$R_T^{\mu\nu} \equiv R^{\mu\nu} - \frac{1}{4}R g^{\mu\nu}$,
to the corresponding traceless
energy-momentum tensor   $T_T^{\mu \nu} $ \cite{Wei89}.
It should be noted that the
energy-momentum conservation law $\nabla_\mu T^{\mu \nu}=0$ does {\it not}
follow from this equation of motion, but has to be imposed separately.
Assuming energy-momentum conservation, it then follows that

\begin{equation}
R+T = - \hat{\Lambda}   \label{trace}
\end{equation}
where $\hat{\Lambda}$ is a constant
(and $\frac{8\pi G}{c^{4}}=1$ and $c=1$). Using the 
contracted Bianchi identity, $\nabla_{\nu}G^{\mu \nu} = 0 $, we then obtain

\begin{displaymath}
  G^{\mu \nu} = T^{\mu \nu} + {{\Lambda}} g^{\mu \nu},
\end{displaymath}
where $\Lambda$ is given in terms of $\hat{{\Lambda}}$ and the vacuum energy
density (part of the energy-momentum tensor) $\rho_{\rm vac}\equiv{\Lambda}_{vac}$.
Hence the cosmological constant $\Lambda$ naturally appears in terms of a constant of
integration in unimodular gravity.

Therefore,  the
theory acquires a new {\em integrability condition\/} \cite{Wei89}.
Both the trace-free FE
and the matter conservation equations are {\em assumed\/}; the integrability condition 
follows from these equations. 
Hence, we obtain the differential relations which 
are functionally equivalent to the full EFE (where the spacetime volume density $\sqrt{g}$ is {\em not\/}
a dynamical variable), where the cosmological constant is thus given in terms of an arbitrary
integration constant $\hat{\Lambda}$ and is not given explicitly by the vacuum energy
$\Lambda_{\rm vac}$.

This is an old proposal essentially initiated by Einstein \cite{Einstein:1915ca}
 and more recently
it has been developed under the name of {\em{unimodular gravity}} 
\cite{NG99,smolin_uni,unruh_uni}.
It has been suggested that unimodular gravity can be used to eliminate problems caused by the nature of the
cosmological constant as well as to resolve the discrepancies between theory and
observation, while not introducing any exotic terms such as quintessence or dark energy
into the analysis of the EFE \cite{ELLIS,NG99,unruh_uni}. Indeed,
although unimodular gravity does not give a unique value for the
effective cosmological constant, it has the potential to solve the
huge discrepancy between theory and observation.
With a suitable high-energy cut-off, the vacuum
energy density is estimated by Weinberg \cite{Wei89} to be of the
order
$\rho_{\rm vac} \simeq 2 \times 10^{71}\,{\rm GeV}^4$,
whereas the effective value of the cosmological constant as
determined by astronomical observations is of the order
$\rho_{\rm obs} \simeq 10^{-47}\,{\rm GeV}^4$.
 However, there is no longer  a
cosmological constant problem.
For example, for a perfect fluid the matter source
term is the manifestly trace-free stress tensor
$\left(\rho+{p}\right)
\left(u_a u_b+(1/4)\,g_{ab}\right)$;
hence, matter enters the FE only in terms of the
inertial mass density $(\rho+p)$, which vanishes in the case of a cosmological constant
(e.g., see \cite{recent}).

Unimodular gravity has also been utilized in the study of the quantization of GR
\cite{unruh_uni,smolin_uni}.
The Hamiltonian of a generally covariant theory
is zero, so in a sense there is no evolution, but
since unimodular gravity is not generally covariant, the classical 
{\it problem of time} is avoided  \cite{unruh_uni}.
In addition, in unimodular gravity quantum gravitational 
factor ordering ambiguities
are alleviated \cite{smolin_uni}.

\section{Averaging Proposal}

We wish to exploit the structure of
unimodular gravity to suggest an alternative approach to averaging
in cosmology.
Within unimodular gravity we need to average the trace-free part of the FE
and the trace of the FE separately.

\noindent
(1) 
{\em{Average trace-free part of the FE}}: Here the resulting correlation 
tensor must consequently be trace-free.
If the form of the resulting equations are of the algebraic form of a `perfect fluid', as in the
cosmological application (with a large scale FLRW geometry), then the correlation 
tensor must be of the form of an effective energy momentum tensor $T^{eff}_{~~ij}$ for which the trace 
$T^{eff}= -{\rho} + 3p = 0$, corresponding to a radiation fluid \cite{wald}. Note that if the matter 
is dust, then $T^{tot} = T^{dust} + T^{eff} = -{\rho_d} -{\rho_r} + 3p_r
= -[{\rho_d} + 2 {\rho_r}] + [{\rho_r} + 3p_r]$, which could be 
(trivially) reinterpreted as
a renormalized dust term (with energy density $[{\rho_d} + 2 {\rho_r}])$ and 
a term corresponding to a constant spatial curvature 
(with $ [{\rho_r} + 3p_r]$) \cite{CP,CPZ}.

\noindent
(2) {\em{Average the trace of the FE}}: In this case we only need to work 
with the (generalized) Friedmann eqn. (\ref{trace}).
{\footnote{Note that the sum of the averaged energy-momentum tensor and the correlation tensor is 
covariantly conserved; the question of whether 
the averaged energy-momentum is separately conserved with respect to the averaged geometry
is determined by averaging the energy-momentum  conservation equation 
(if it is not, then there is an effective
interation between the  averaged energy-momentum tensor and the 
correlation tensor \cite{Bill}).}}

The problem of averaging is then effectively reduced to considering 
{\em{the average of a single scalar}} eqn.
(see \cite{Scalar}).

\section{Example: Lema\^{\i}tre-Tolman-Bondi model}

The exact spherically
symmetric dust Lema\^{\i}tre-Tolman-Bondi
(LTB) model \cite{LTB}, which can be regarded as an
exact inhomogeneous generalization of the FLRW solution, 
can be rewritten in VPC 
$(t,x,u,\phi)$ \cite{CP}. Taking $A=A(t,x)$, the line-element becomes
\begin{equation}
ds^2=-\left(1-\frac{U^2}{A^4}\right)dt^2-2\frac{U}{A^4}dtdx+\frac{dx^2}{A^4}+A^2\left[\frac{du
^2}{1-u^2}
+(1-u^2)d\phi^2\right],     \label{ltbvpc}
\end{equation}
which has $det(g)=1$ as desired, where $U(t,x)$ is defined as
\begin{equation}
U(t,x)=-\frac{2A_{t}A_{x}+AA_{tx}}{2A_{x}^{2}+AA_{xx}}.
\label{conu}
\end{equation}
The constraints on the original LTB metric ensuring a dust solution 
are given in \cite{CP}.
For general functions $A$ and $U$, the Ricci scalar of the metric (\ref{ltbvpc}) is
given by

\begin{eqnarray}
  R = \frac{2}{A^2}  \left( 1 - 5 A_x ^2 A^4 + 3U^2 A_x^2
-2 U_x A A_t - 2 A_{xx}   A^5 + 3 A_t^2 + 6 U A_x A_t  \right. \nonumber \\ \left.
-2 A U U_x A_x + A^2 U U_{xx} + A^2 U_{tx} + A^2 U_x^2
\right) 
\label{riccisc}
\end{eqnarray}

The spatially flat ($E_{0}=0$) \frw model in VPC is 
given by the metric (\ref{ltbvpc}) with

\begin{eqnarray}
A(t,x) \equiv A_0=(3x)^{1/3}, &\  & U(t,x)=\frac{2x}{t-t_{B}},
\label{frwflat}
\end{eqnarray}
where, strictly speaking, the degenerate form for $U(t,x)$
does not follow
directly from eqn. (\ref{conu}) (however, eqn. (\ref{riccisc})
is valid for (\ref{frwflat}) and $R \sim (t-t_{B})^{-2}$). 
Defining $A_0 = r S(t)$, the Ricci scalar of the FLRW metric with positive 
curvature constant $k=+1$
is given by
$R=6 [{S} {S_{tt}}  + {S_{t}^{~2}} + {k}]{S^{-2}}$.
For the zero-curvature Einstein de-Sitter metric,
$S \sim t^{\frac{2}{3}}$, and setting $t_{B}=0$, we get the approximate expression: 
\begin{equation}
R= \frac{4}{3} t^{-2} [1 + \frac{9}{2}kt^{\frac{2}{3}} + {\mathcal {O}}(t)],\label{feds}
\end{equation}
consistent with the expression given in \cite{CP} 
(with $E_0 = \frac{9}{2}k \equiv c r_0^{~2}$).

\subsubsection{A Perturbative Solution}

let us assume that $t_{B}(r)$ is zero, which implies that the bang time is uniform and
we are consequently restricting our choice of LTB models to those with no decaying modes.
We shall also consider solutions of
the LTB metric in VPC as perturbations about the spatially flat \frw model given by
(\ref{frwflat}).  In this respect our approximate solution will be an expansion with
respect to $E_{0}$ and we require the Einstein tensor to have the form of dust (after truncation of terms of ${\mathcal O}(E_{0}^{2})$ or
higher).  We begin by making the formal expansion for  $ A$ in the form: \begin{equation}
A(t,x)=A_{0}+\alpha_{1}x^a t^b E_0 + \alpha_2 x^c t^d E_{0}^2, \label{Ransatz}
\end{equation} 
\noindent where $\alpha_1$, $\alpha_2$, $a$, $b$, $c$ and $d$ are constants.  
We can use eqns.  $(\ref{Ransatz})$ and $(\ref{conu})$ to obtain $U(t,x)$.  Calculating
the Einstein tensor and requiring it have the form of dust (up to
order $E_0^2$) allows us to
determine the constants in our perturbative solution (we obtain:
$a=1/3$, $b=0$, $c=5/6$ and $d=-1$ \cite{CP}).

The expression that results from substituting $U$ in terms of $A$
using equation (\ref{conu}) and the expression (\ref{Ransatz}) for $A$ (with 
the given powers of $x$
and $t$ in our particular perturbative solution) leads to the expression for the Ricci scalar $R$
(keeping only terms up to ${\mathcal {O}}(E_0^2)$):

\begin{equation}
  R = \frac{4}{3t^2} - 4 E_0 \alpha_1 x^{-2/3} - 
E_0^2 \left( 15 \alpha_1 x^{-1/6} t^{-1} + 2 \alpha_2^2 3^{-1/3} x^{-2/3}  \right).
\end{equation}
Defining $r^3 = 3x~t^{-2}$, we obtain

\begin{equation}
  R = \frac{4}{3t^2} + a E_0 \alpha_1 r^{-2} t^{-4/3} + 
E_0^2 \left(b \alpha_1 r^{-1/2} t^{-4/3} 
+ c \alpha_2^2 r^{-2} t^{-4/3}  \right) \label{ricci}
\end{equation}
where
\begin{equation}
a \equiv - 4\times3^{2/3},~b \equiv -15\times3^{1/6},~c \equiv - 2\times3^{1/3}.
\end{equation}

Finally, we obtain the averaged version of the Ricci scalar equation
by integrating eqn. (\ref{ricci}) over the radial variable $r$, where $r_0$ is the (radial)
averaging length scale:

\begin{equation}
  R = \frac{4}{3t^2}\left(1 + {\bar{a}} E_0 \alpha_1 t^{2/3} + 
E_0^2 \left[{\bar{b}} \alpha_1 t^{2/3} 
+ {\bar{c}} \alpha_2^2 t^{2/3}  \right] \right)
\end{equation}
(where the `barred' constants are the appropriately $r_0$-renormalized constants).
We see that {\it{all of the correction terms}} (correlations) introduced
by {\it{averaging the Ricci scalar equation}}
are of the form of a {\it{spatial curvature term}}
(\ref{feds}), 
which is
consistent with the results of \cite{CP}. 
{\footnote{We note that for this perturbative solution, we obtain higher order correction
terms of the form $\sim t^{-2}$, which can be interpreted as a renormalization of 
$A_0$ in the exact dust solution.  We
also note that, in principle, for the second order terms (${\mathcal O}(E_{0}^2)$)
to be  formally comparable, 
${\alpha_{1}}^{2} \sim  \alpha_{2} {r_0}^{\frac{3}{2}}$.}}

\section{Discussion}

Recent observations are usually interpreted as implying that the
Universe  is  very  nearly  flat,  currently   accelerating  and
indicating   the  existence  of  dark  matter  and  dark  energy
\cite{SN}.  A cosmological constant is a candidate  for the dark  energy.
Averaging can have
a very significant dynamical effect on the evolution of the Universe; the correction terms change the
interpretation of observations so that they need to be accounted for carefully to determine if the models may be
consistent with an accelerating Universe. 
Indeed, it has been argued that a  more   conservative   approach  to  explain  the
acceleration  of the  Universe  without the introduction  of exotic
fields   might be  to   utilize   a   backreaction   effect   due   to
inhomogeneities of the Universe.

In this paper we have argued that a rigorous approach to cosmological averaging
(and necessary for studying cosmological data)
is perhaps most naturally studied within the context of unimodular gravity.
In the simple example studied here, 
we found  that all correction terms introduce correlations 
of the form of a spatial curvature term \cite{CP}.

As another simple illustration, in the special case ($C_{\epsilon} = 0, {\epsilon}_i = 0$) 
of the exact solution representing a 
two-scale Buchert average of the EFE for an
inhomogeneous universe approximating the observed Universe
\cite{Wiltshire}, we have that 
$S(t) = \alpha t^{2/3} [1 + \beta t]^{1/3}$, where 
$\alpha \equiv a_0 (3 H_0)^{2/3} (1 - f_{v0})^{1/3} (2 + f_{v0})^{-2/3}$ and
$\beta \equiv 3 f_{v0} H_0 (1 - f_{v0})^{-1}(2 + f_{v0})^{-1}$, and the physical constant
$f_{v0}$ is related to the 
scales representing the voids and the bubble walls surrounding them 
(within which clusters of galaxies are
located). In this example, the Ricci scalar is again of the form of eqn.
(\ref{feds}).

In future work we intend to consider this averaging scheme in 
more general cosmological contexts. In particular, we wish to study
approximate solutions within linear perturbation theory. A first step
will be to
develop perturbation theory within unimodular gravity
\cite{Backreaction}.

{\em Acknowledgements}.
 This work was supported, in
part, by NSERC.

\end{document}